\documentclass{article}
\usepackage{spconf,amsmath,graphicx}
\usepackage{enumitem}
\usepackage{booktabs}

\title{A practical framework for multi-domain speech recognition
and an instance sampling method to neural language modeling
}
\name{Yike Zhang, Xiaobing Feng, Yi Liu, Songjun Cao, Long Ma}
\address{Tencent Cloud Xiaowei, Beijing, China}

\begin{document}
%\ninept
%
\maketitle
\begin{abstract}
    Automatic speech recognition (ASR) systems used on smart phones or vehicles
    are usually required to process speech queries from very different domains.
    In such situations,
    a vanilla ASR system usually fails to perform well on every domain.
    This paper proposes a multi-domain ASR framework for Tencent Map,
    a navigation app used on smart phones and in-vehicle infotainment systems.
    The proposed framework consists of three core parts:
    a basic ASR module to generate n-best lists of a speech query,
    a text classification module to determine which domain the speech query belongs to,
    and a reranking module to rescore n-best lists using domain-specific language models.
    In addition,
    an instance sampling based method to training neural network language models (NNLMs)
    is proposed to address the data imbalance problem in multi-domain ASR.
    In experiments,
    the proposed framework was evaluated on navigation domain and music domain,
    since navigating and playing music are two main features of Tencent Map.
    Compared to a general ASR system,
    the proposed framework achieves a relative 13\% $\sim$ 22\% character error rate reduction 
    on several test sets collected from Tencent Map and our in-car voice assistant.
\end{abstract}
\begin{keywords}
multi-domain ASR, 
multi-domain NNLM,
instance sampling, 
text classification, 
logistic regression
\end{keywords}

\section{Introduction}
\label{sec:intro}
Smart phones have become a prominent feature in the everyday work 
and personal life of the general public.
Increasing apps support voice interaction.
Since functions of apps vary from one another,
automatic speech recognition (ASR) systems used for these apps 
should be optimized for corresponding domains.
Nowadays,
more and more functions are packed into a single app.
As a result,
ASR systems used for some apps are required to 
recognize speech queries from more than one domains.
A general ASR system 
that supports recognizing queries from multiple common domains 
usually fails to perform well on specific domains.
Intuitively, 
a multi-domain ASR system, 
which has been optimized for multiple specific domains, can handle this problem.
However, 
it is difficult to directly build such a multi-domain ASR system since
linguistic knowledge varies from domains to domains. 

This paper focuses on building a multi-domain ASR system for our real-time navigating app, Tencent Map. 
Tencent Map has two main features,
navigating and playing songs.
Therefore, 
our speech recognition system should support point of interest (POI) search and music search.
There are very big differences in these two domains.
When searching for places,
people are more concerned about massive long-tailed POI names.
In contrast,
when searching for music, 
people prefer a limited number of songs or singers
which have been popular lately.
% In consideration of the features of POI and music,
% main difficulties to build our multi-domain ASR system are as follows.
Based on the above analysis,
the main difficulties in our case are the recognition of long-tailed POI names and the timely update of music knowledge.
In addition,
some songs and POI names are similar in pronunciation. 
For such cases,
it is difficult to get the right result without domain information.
% There are also some homophone POI names in different areas.
Moreover,
data imbalance is also a problem in our case
% there are imbalance among the text corpus from different domains since
since there are far more POI names than hot songs and singers.
In order to deal with the above problems,
this paper proposed a novel framework for multi-domain ASR.
The main contributions of this paper are as follows:
\begin{itemize}
    \item 
    This paper proposed a multi-domain ASR framework 
    which consists of a basic ASR module, 
    a text classification module,
    and a reranking module.
    Concretely,
    the text classification module determines which domain a query belongs to 
    using the top one hypothesis in n-best lists 
    of the basic ASR module output.
    Then,
    the reranking module rescores n-best lists using domain-specific language models.
    In such a way,
    the linguistic knowledge in navigation domain can be distinguished from that in music domain
    % In addition,
    % instead of ground-truth domain information, 
    % using predicted domain information from the text classification module 
    % makes the proposed framework suitable for more practical application scenarios.
    and the linguistic knowledge in a specific domain can be easily updated without affecting the ASR performance on other domains.
    In order to further improve the recognition accuracy of long-tailed POI names,
    % and 
    % alleviate the homophone problem for POI names,
    geographical language models proposed in our previous work \cite{cao2021improving} are also integrated into the reranking module.
    
    \item 
    An instance sampling based approach is proposed to deal with the data imbalance problem in neural network language model (NNLM) training.
    The proposed approach samples training examples from corpus in different domains according to pre-calculated probabilities.
    The core idea of the proposed approach is that
    the sample size for a certain domain is in proportion to the importance of the corpus in that domain.
    Therefore, 
    the proposed approach can avoid NNLMs performing unsatisfactorily on the domains 
    which are important but with insufficient training corpus.
    
\end{itemize}

The rest of this paper is organized as follows.
Section \ref{sec:priorwork} describes prior researches related to this work.
The proposed multi-domain ASR framework and instance sampling based approach to training NNLMs 
are presented in Section \ref{sec:mdasr} and Section \ref{sec:mdnnlm} respectively.
Experiments and results are shown in Section \ref{sec:expriments}.
Finally, 
Section \ref{sec:conclusions} concludes this paper.

%%%%%%%%%%%%%%%%%%%%%%%%%%%%%%%%%%%%%%%%%%%%%%%%%%%%%%%%%%%%%%%%%%%%%%%%%%%%%%%%%
\section{Prior Work}
\label{sec:priorwork}
This paper mainly focuses on domains in the linguistic scene,
which is related more to language modeling.
Most previous researches on multi-domain speech recognition or multi-domain language modeling 
focus on domain adaptation techniques 
\cite{alumae2013multi}, 
\cite{tilk2014multi}, 
\cite{chen2015recurrent}, 
\cite{tuske2016investigation}, 
\cite{deena2018recurrent}, 
\cite{hentschel2019feature}.
These adaptation techniques, 
which are only suitable for neural network models,
fall into two categories: 
feature based approaches and model based approaches.
In feature based approaches,
unsupervised features, 
such as latent Dirichlet allocation (LDA) \cite{blei2003latent},
probabilistic latent semantic analysis (PLSA) \cite{hofmann1999probabilistic},
or hierarchical Dirichlet process (HDP) \cite{teh2006hierarchical},
are used as auxiliary features to represent implicit domain information
\cite{chen2015recurrent}, 
\cite{hentschel2019feature}.
In addition,
one hot encoding of ground-truth domain information is also used when it is available during inference
\cite{alumae2013multi}, 
\cite{tilk2014multi}, 
\cite{hentschel2019feature}.
As for model based approaches,
extra components,
such as linear hidden network (LHN) \cite{gemello2007linear},
learning hidden unit contributions (LHUC) 
\cite{swietojanski2014learning}, 
\cite{samarakoon2016subspace},
or adapters \cite{lee2020adaptable},
are added to basic models to learn domain-specific information
\cite{hentschel2019feature}, 
\cite{lee2020adaptable}.
Unfortunately,
these approaches cannot be directly applied to our case
since new songs may release at any time.
As a result,
NNLMs should be updated frequently.
This takes up a lot of computing resources.
In addition,
updating a multi-domain NNLM only on a single domain may lead to performance decrease on other domains 
since there are plenty of parameters shared among different domains.

In addition,
the above approaches do not consider the problem of data imbalance.
Prior researches address data imbalance either at data level 
\cite{wunsch2009instance},
\cite{crone2012instance},
\cite{buda2018systematic}
or at classifier level\cite{buda2018systematic}, 
\cite{li2020overcoming}.
Oversampling and undersampling are most common data level approaches.
Classifier level approaches usually modify original output class probabilities.
Language modeling is not a typical classification problem.
Therefore, 
data level approaches are more suitable for our case.
However,
there has been little research on the sampling methods for language modeling tasks.

%%%%%%%%%%%%%%%%%%%%%%%%%%%%%%%%%%%%%%%%%%%%%%%%%%%%%%%%%%%%%%%%%%%%%%%%%%%%%%%%%
\section{Multi-domain ASR}
\label{sec:mdasr}
\begin{figure}[t]
  \centering
  \includegraphics[width=\linewidth]{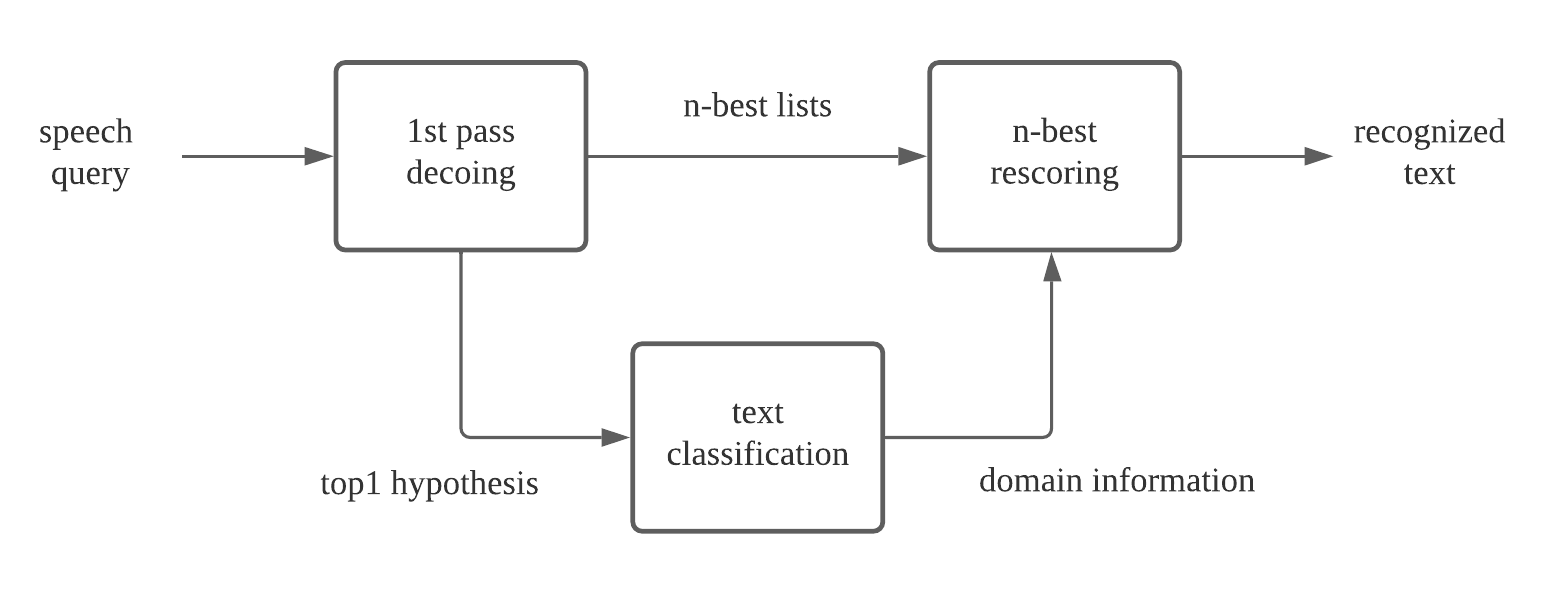}
  \caption{Schematic diagram of the proposed multi-domain ASR framework.}
  \label{fig:mdasr}
\end{figure}

The proposed multi-domain ASR framework consists of three main modules:
a basic ASR module to conduct first-pass decoding and generate top N hypotheses of a speech query,
a text classification module to determine which domain the speech query belongs to,
and a reranking module to rescore n-best lists of the first-pass decoding output 
using domain-specific language models.
Figure \ref{fig:mdasr} shows the diagram of the proposed multi-domain ASR framework.

The basic ASR module consists of a general acoustic model and a multi-domain n-gram language model.
In our implementation, 
the multi-domain n-gram model was simply built by 
interpolating n-gram models in the domain of navigation, music, video, news, fiction, and chat. 
The interpolation weights were computed on a development set.

Text classification has been widely studied and addressed in many real applications over the last few decades\cite{kowsari2019text}.
Recently,
more and more advanced methods have been proposed.
Logistic regression (LR) is a classic text classification algorithm \cite{genkin2007large}, 
\cite{ifrim2008fast},
which is simple but effective.
Since this paper does not focus on the optimization of text classification algorithms
and low latency is preferred in our case,
we adopt LR models as the text classification module in the proposed framework.
Formally,
a LR model with parameter $\theta$ divides a sentence $s$ into a category $C$ 
with the probability 
\begin{equation}
  P(y = C | s; \theta) = \frac{1}{1 + e^{-\theta^{T} x}}
  \label{eq:lr}
\end{equation}
where
$x$ is a tf-idf \cite{ramos2003using}, 
\cite{aizawa2003information}, 
\cite{zhang2011comparative} vector derived from uni-grams, bi-grams, and tri-grams in $s$.
One LR model is built for each domain.
In this work,
we built two LR models. 
One is for navigation domain and the other is for music domain.
At inference,
the predicted domain $y$ of a speech query is 
\begin{equation}
  y \! = \! \mathop{\arg \! \max}_{C_i} \{ P(C_0),P(C_1|h_1;\theta_1), \! \cdots \!, P(C_n|h_1;\theta_n) \}
  \label{eq:class}
\end{equation}
where
$h_1$ is the top one hypothesis generated by the ASR module,
$n$ is the number of domains,
$C_0$ represents a special domain ``other'',
and $P(C_0)$ is a constant representing the minimum threshold of predicted probabilities.
Specifically,
a speech query is divided into ``other'' domain $C_0$
only if the outputs of all LR models are less than $P(C_0)$.

The reranking module contains multiple language models.
A multi-domain NNLM was trained with the proposed approach in Section \ref{sec:mdnnlm}.
For each domain,
a domain-specific n-gram model was trained using only in-domain corpus.
For the ``other'' domain,
a general n-gram model was trained using corpus from all domains.
In order to improve the performance of the proposed framework on navigation domain,
a group of geographical n-gram models are used.
Each geographical n-gram model was trained with POI names in the corresponding geographical area.
More details about geographical models can be found in our previous work \cite{cao2021improving}.
In the reranking stage,
a new language score 
\begin{equation}
P^{r}(h_i) \! = \!
\left\{
    \begin{array}{ll}
    \!\!\! \alpha P_{r}(h_i) + (1 - \alpha) P_{m}(h_i),              \!\!\! & y \! = \! C_{m} \\
    \!\!\! \alpha P_{r}(h_i) + \beta P_{n}(h_i) + \gamma P_{g}(h_i), \!\!\! & y \! = \! C_{n} \\
    \!\!\! \alpha P_{r}(h_i) + (1 - \alpha) P_{o}(h_i),              \!\!\! & y \! = \! C_{o} \\
    \end{array}
\right.
\label{eq:lm}
\end{equation}
%
% $P^{r}(h_i)$
is computed for each hypothesis $h_i$ in n-best lists,
where $P_{r}(h_i)$, $P_{m}(h_i)$, $P_{n}(h_i)$, $P_{o}(h_i)$, $P_{g}(h_i)$
represent probabilities derived from the multi-domain NNLM,
the n-gram model built for music domain,
the n-gram model built for navigation domain,
the general n-gram model built for the ``other'' domain,
and a specific geographical n-gram model respectively.
Constants $\alpha$, $\beta$, $\gamma$ control the contributions of different models,
and $\gamma = 1 - \alpha - \beta$.
Conditions $y = C_{m}$, $y = C_{n}$, y = $C_{o}$
represent the text classification module divides the query into 
the domain of music, navigation, and ``other'' respectively.
By introducing domain-specific n-grams models into the reranking module,
the linguistic knowledge in a specific domain can be easily updated.
Finally,
n-best lists are reranked according to the score %$S(h_i)$ 
\begin{equation}
    S(h_i) = \eta P^{a}(h_i) + (1 - \eta) ( \mu P^{l}(h_i) + (1 - \mu) P^{r}(h_i)) 
    \label{eq:rerank}
\end{equation}
where $P^{a}(h_i)$, $P^{l}(h_i)$ are the acoustic score and language score derived from the ASR module.
$\eta$ controls the contributions of the acoustic model and language models,
$\mu$ controls the contributions of different language models.

%%%%%%%%%%%%%%%%%%%%%%%%%%%%%%%%%%%%%%%%%%%%%%%%%%%%%%%%%%%%%%%%%%%%%%%%%%%%%%%%%
\section{Instance sampling}
\label{sec:mdnnlm}
Instance sampling \cite{wunsch2009instance}, 
\cite{crone2012instance} is an efficient method to addressing data imbalance.
This paper studies how to apply instance sampling to neural network language modeling.
In order to make full use of the core corpus,
this paper adopts an oversampling based approach.
The core part of sampling based approaches is to determine sampling size.
In our case,
the core problem is how to determine sampling size for each domain.

Steps of the proposed approach are listed as follows:
\begin{enumerate}[label= \arabic*), itemindent=1em]
    \item 
    Dividing all corpus into $D$ categories by their sources 
    so that corpus in one category is of the same domain.
    
    \item
    Counting the number of sentences in each category, $s_1, s_2, \cdots, s_D$.
    
    \item 
    Estimating an n-gram model for each category from corpus only in the corresponding category.
    
    \item
    Computing interpolation weights 
    $w_1, w_2, \cdots, w_D$  ($\sum_{i} w_i = 1$) of the n-gram models on a development set. 
    These weights represent the importance of the corpus in different categories.
    The larger the weight is,
    the more important the corpus in the corresponding category is.
    
    \item
    Determining the total number of sentences to be sampled or the sampling size
    \begin{equation}
        N = \frac{\max\{ s_1, s_2, \cdots, s_K \}} 
                 {w_{\mathop{\arg \max}_{i} \{ s_1, s_2, \cdots, s_K \}}},
        \label{eq:samplingN}
    \end{equation}
    where $K$ ($K \le D$) is the number of core categories,
    the numerator is the number of sentences in the category which has the most sentences in the core categories,
    and the denominator represents the interpolation weight of the corresponding category.
    In this work,
    there are six categories in total including 
    navigation, music, video, news, fiction, and chat.
    The core categories are navigation and music.
    Therefore, 
    we set $N=6$ and $K=2$.
    In general case,
    core categories can be determined by 
    % selecting the categories with top $K$ largest interpolation weights. 
    selecting the top K categories with the largest interpolation weights. 
    In this way,
    we can make full use of all the core corpus.
    
    \item
    Determining the number of sentences to be sampled from each category by letting 
    \begin{equation}
        \begin{aligned}
        d_1: d_2: \cdots: d_D & = w_1: w_2: \cdots: w_D, \\
                              s.t. \sum\nolimits_{i} d_i & = N
        \end{aligned}
        \label{eq:samplingd1}
    \end{equation}
    where 
    \begin{equation}
        d_i = N w_i
        \label{eq:samplingd2}
    \end{equation}
    is the number of sentences to be sampled from category $i$.
    
    \item
    Sampling each sentence from category $i$ with probability
    \begin{equation}
        r_i =
        \left\{
            \begin{array}{ll}
            \min \{ d_i / s_i, 0.95 \},     & d_i  <   s_i \\
            \min \{ d_i / (ms_i), 0.95 \},   & d_i \ge s_i \\
            \end{array}
        \right. ,
        \label{eq:samplingr1}
    \end{equation}
    where 
    \begin{equation}
        m =  \lceil d_i / s_i \rceil
        \label{eq:samplingr2}
    \end{equation}
    means rounding up the value $d_i / s_i$ to an integer.
    Eq.(\ref{eq:samplingr1}) and Eq.(\ref{eq:samplingr2}) imply that 
    each sentence in category $i$ should be repeated for $m$ times before sampling
    if the number of sentence to be sampled $d_i$ is greater than the number of sentence $s_i$ in category $i$.
    In addition,
    we limit the sampling probability $r_i$ no more than 0.95 to filter garbage sentences 
    since a large proportion of corpus is from our online ASR engines.
    An intuitive explanation of Eq.(\ref{eq:samplingr1}) is that 
    a sentence in a more important category is more likely to be picked
    since $r_i$ is proportional to $d_i$, 
    and therefore is proportional to $w_i$.
    
    \item
    Generating the training set $S$ by mixing and shuffling all sentences sampled from different categories.

\end{enumerate}

After getting the training set $S$,
we can train a multi-domain NNLM immediately.

%%%%%%%%%%%%%%%%%%%%%%%%%%%%%%%%%%%%%%%%%%%%%%%%%%%%%%%%%%%%%%%%%%%%%%%%%%%%%%%%%
\section{Experiments}
\label{sec:expriments}

\subsection{Setups}
All the speech corpus used to train the acoustic model in the ASR module
is collected from our productions including Tencent Map. 
The text corpus used in this work is collected from our productions 
including Tencent Map, in-car voice assistants, and smart speakers,
as well as the Internet.
The text corpus is divided into six domains,
including navigation, music, video, news, fiction, and chat,
according to their sources.
The sizes of corpus in the corresponding domains %navigation, music, video, news, fiction, and chat domains
are 53GB, 9GB, 7GB, 45GB, 200GB, and 8GB respectively.

In the ASR module,
the acoustic model, 
consisting of two CNN layers and three TDNN-OPGRU \cite{cheng2018output} blocks,
was trained with 40-dimensional PNCC \cite{kim2016power} 
and lattice-free maximum mutual information (LF-MMI) \cite{povey2016purely} criterion. 
The 5-gram language model in the ASR module was trained with all above text corpus.

The text classification module contains two LR models.
One is for navigation domain and the other is for music domain.
The LR models were trained with manually labeled data in corresponding domains.

In the reranking module,
% three 5-gram models were estimated from the corpus in music domain, 
% the corpus in navigation domain,
% and all the text corpus respectively.
three 5-gram models were built for navigation domain, music domain, and ``other'' domain respectively.
The model for navigation/music domain was estimated from the corpus only in the navigation/music domain. 
The model for ``other'' domain was trained using all corpus in the six domains. 
In addition,
for each province,
a 5-gram geographical model was estimated from POI names in the corresponding province.
The POI names used to train geographical models are all from the corpus in navigation domain.
With the proposed instance sampling based approach,
about 400 GB text was sampled from all corpus in different domains to train a multi-domain NNLM.
In order to balance performance and efficiency,
the multi-domain NNLM adopted the architecture of QRNN \cite{bradbury2016quasi}. 
The QRNN model also used an adaptive softmax output layer \cite{joulin2017efficient} to further improve efficiency.

In experiments,
a development set consisting of 10,000 utterances is used to compute interpolation weights in the proposed method to training multi-domain NNLMs.
Four test sets are used to evaluate the proposed mutli-domain ASR framework,
each of which consists of 2,000 utterances.
Both the development set and test sets are collected from our productions.
Specifically,
two test sets, 
POI-map and music-map,
are collected from Tencent Map.
The other two test sets,
POI-car and music-car, 
are collected from our in-car assistant.
POI-map and POI-car only contain utterances in navigation domain,
while music-map and music-car only contain utterances in music domain.

\subsection{Results}
The proposed multi-domain ASR framework is rather complex 
since it contains three core modules and each module consists of multiple components.
In experiments,
we evaluated the proposed framework as detailed as possible.
Table \ref{tab:asr} show the performance of the proposed multi-domain ASR framework
with different configurations on the test sets.

\begin{table*}[th]
  \caption{Character error rates (\%) of the proposed multi-domain ASR framework with different configurations.
  %on four test sets collection from our productions.
  %   
  Classifier indicates domain information was either from LR models or ground-truth.
  %show either LR models or the text classification model.
  %
  The NNLM was either trained by simply mixing corpus from all domains (mixed) or
  with the proposed instance sampling based approach (sampled).
  %``mixed'' means the NNLM is trained by simply mixing corpus from all domains.
  %``sampled'' means the NNLM is trained with the proposed instance sampling based approach.
  %
  DLM represents the domain-specific n-gram models 
  in the domain of music ($L_m$), navigation ($L_n$), and ``other'' ($L_o$).
  %where $L_m,L_n,L_o$ represent the model for music, navigation, and ``other'' domain respectively.
  %
%   GLM indicates whether to adopt geographical models.
  GLM meas geographical language models.
  %DLM and GLM represent domain-specific n-gram models and geographical models.used in the reranking module respectively.
  %
%   ``-'' means corresponding components were not adopted, 
%   while ``Y'' means corresponding components were adopted.
  ``Y''/``-'' means corresponding components were adopted/not adopted.
  }
  \label{tab:asr}
  \centering
  \begin{tabular}{ c|cccc|cccc }
    \toprule
    Configuration  &     classifier&      NNLM&            DLM&    GLM&     POI-map& POI-car& music-map& music-car \\
    \midrule
    C1&              -&         -&              -&      -&        6.54& 2.84& 14.75& 7.63\\
    C2&              -&     mixed&          $L_o$&      -&        6.35& 2.73& 12.29& 6.53\\
    C3&              -&     mixed&          $L_o$&      Y&        6.25& 2.53& 17.11& 8.44\\
    C4&             LR&     mixed&  $L_m/L_n/L_o$&      Y&        5.67& 2.43& 12.31& 6.59\\
    C5&   ground-truth&     mixed&  $L_m/L_n/L_o$&      Y&        5.61& 2.36& 11.79& 6.48\\
    C6&              -&   sampled&          $L_o$&      -&        6.19& 2.53& 11.85& 5.95\\
    C7&             LR&   sampled&  $L_m/L_n/L_o$&      Y&        5.67& 2.38& 11.66& 5.93\\
    \bottomrule
  \end{tabular}
\end{table*}

% \begin{table*}[th]
%   \caption{The character error rate of the proposed multi-domain ASR system with different configurations on four test sets collection from our productions.}
%   \label{tab:asr}
%   \centering
%   \begin{tabular}{ clcccc }
%     \toprule
%     No & 
%         \multicolumn{1}{c}{Configurations}& 
%         Navigation-TM& 
%         Navigation-ICA& 
%         Music-TM& 
%         Music-ICA \\
%     \midrule
%     C1& basic ASR module&
%         6.54& 2.84& 14.75& 7.63\\
%     C2& %\quad 
%         C1 + reranking&
%         6.35& 2.73& 12.29& 6.53\\
%     C3& %\quad \quad 
%         C2 + geographical models&  
%         6.25& 2.53& 17.11& 8.44\\
%     C4& %\quad \quad \quad 
%         C3 + text classification&
%         5.67& 2.43& 12.31& 6.59\\
%     C5& %\quad \quad \quad 
%         C3 + oracle text classification&
%         5.61& 2.36& 11.79& 6.48\\
%     C6& %\quad \quad 
%         C2 + multi-domain QRNN&
%         6.19& 2.53& 11.85& 5.95\\
%     C7& %\quad \quad \quad \quad  
%         C4 + multi-domain QRNN& 
%         5.67& 2.38& 11.66& 5.93\\
%     \bottomrule
%   \end{tabular}
% \end{table*}

The performance of the basic ASR module in the proposed framework was evaluated first. 
Character error rates (CERs) on different test sets are provided in the first row (C1).
Generally,
the basic ASR module performs better on navigation domain than on music domain.
% since it focuses on navigation domain from the very beginning.
Rescoring n-best lists of the first-pass decoding output can usually achieve lower CERs.
In C2,
the n-best lists of the basic ASR module output were rescored 
with the general 5-gram model built for ``other'' domain 
and a QRNN model trained by simply mixing corpus from all domains. 
Namely, 
the n-best lists were resored with the last formula in Eq.(\ref{eq:lm}).
As results shows,
more than 10\% relative CER reduction is achieved on music domain.
Our previous work \cite{cao2021improving} indicates that 
geographical language models can significantly improve the recognition accuracy of POI names.
However,
when geographical models were directly applied in the reranking module 
by replacing $P_{n}(h_i)$ in the second formula of Eq.(\ref{eq:lm}) with $P_{o}(h_i)$,
only slight improvement is achieved on navigation domain (C3).
Furthermore,
the performance on music domain is seriously deteriorated.
This indicates that
optimizing one domain of a general ASR system may lead to performance degradation on other domains.
With the help of the text classification module,
domain-specific reranking of Eq.(\ref{eq:lm}) can be achieved.
Under configuration C4,
about 15\% relative CER reduction is reached on navigation domain 
and there is no performance degradation on music domain.

In order to investigate why no improvements on music domain after using text classification,
more experiments on the text classification module were conducted.
Table \ref{tab:nlu1} and Table \ref{tab:nlu2} show 
the precision, recall, and F1 measure of the text classification module 
on different domains.
\begin{table}[th]
  \caption{Performance of the text classification module with top 1 hypotheses as input.}
  \label{tab:nlu1}
  \centering
  \begin{tabular}{ lccc }
    \toprule
    Domain&     Precision& Recall& F1-measure\\
    \midrule
    Navigation& 0.9191& 0.8814& 0.8999\\
    Music&      0.9737& 0.7695& 0.8597\\
    Other&      0.3829& 0.7920& 0.5162\\
    \bottomrule
  \end{tabular}
\end{table}

\begin{table}[th]
  \caption{Performance of the text classification module with references as input.}
  \label{tab:nlu2}
  \centering
  \begin{tabular}{ lccc }
    \toprule
    Domain&     Precision& Recall& F1-measure\\
    \midrule
    Navigation& 0.9603& 0.8915& 0.9246\\
    Music&      0.9754& 0.8628& 0.9156\\
    Other&      0.4553& 0.7998& 0.5803\\
    \bottomrule
  \end{tabular}
\end{table}

When references of the speech queries are used as the inputs of the text classification module,
%precision and F1 measure on navigation and music domains are greater than 90\%, and recall is greater than 85\% on both domains.
satisfactory results are reached on both navigation domain and music domain.
However,
when top one hypotheses in the n-best lists of the ASR module output 
are used as the inputs of the text classification module, 
the recall on music domain decreases significantly. 
Consequently,
in C4,
some queries in music domain were mistakenly divided into the ``other'' domain and 
a mismatched n-gram model was used to rescore corresponding n-best lists.
This may cause performance degradation in music domain.
In order to find out the performance limit of the text classification module in the proposed multi-domain ASR framework,
ground-truth domain labels, 
instead of predicted domain labels, 
were used to select domain-specific n-gram models in the reranking module (C5).
Results are shown in the fifth row of Table \ref{tab:asr}.
Compared to using predicted domain labels,
less than 5\% CER reduction is achieved on all test sets by using ground-truth domain labels.
This indicates that optimizing the text classification module can further
improve the performance of the proposed multi-domain ASR framework to some extent.

Finally,
the multi-domain QRNN trained with the proposed instance sampling based method was evaluated .
Results are presented in the last two rows of Table \ref{tab:asr} (C6 and C7).
Results show that 
the proposed instance sampling based method can significantly reduce the CER on music domain,
which further demonstrates the proposed method can effectively alleviate the data imbalance problem.

%%%%%%%%%%%%%%%%%%%%%%%%%%%%%%%%%%%%%%%%%%%%%%%%%%%%%%%%%%%%%%%%%%%%%%%%%%%%%%%%%
\section{Conclusions}
\label{sec:conclusions}
This paper proposes a multi-domain ASR framework for Tencent Map,
consisting of a basic ASR module, a text classification module, and a reranking module.
In the proposed framework,
the linguistic knowledge in a certain domain can be easily updated without affecting the ASR performance other domains 
and domain-specific rescoring can be achieved.
% The proposed framework have two highlights:
% 1) a text classification module is integrated into the proposed system 
% so that domain related reranking of n-best lists can be achieved in multi-domain ASR;
% 2) an instance sampling approach is proposed to deal with the data imbalance problem
% in NNLM training,
% which can effectively improve the performance of NNLMs on under-resourced domains.
Besides,
an instance sampling based approach to training NNLMs is proposed to address the data imbalance problem in multi-domain ASR.

Experiments show 
the proposed framework achieves more than 10\% CER reduction on navigation domain
and more than 20\% CER reduction on music domain, 
while a vanilla ASR system fails to simultaneously improve recognition accuracies on both domains.
Further analysis on the text classification module demonstrates that 
lower CER results can be reached by optimizing the text classification algorithm.
Results show 
the NNLM trained with the proposed instance sampling based method reaches a 5\% $\sim$ 10\% relative CER reduction on  music domain,
which hints the proposed instance sampling based method can efficiently alleviate the data imbalance problem.

\bibliographystyle{IEEEbib}
\bibliography{strings,refs}

\end{document}